\documentclass[a4paper,11pt]{article}
\pdfoutput=1
\usepackage{jheppub} 
\usepackage{amsmath,amstext,amssymb}
\usepackage{comment,url}
\usepackage{graphicx,subfig}
\newcommand{\nn}{\nonumber\\}

\title{Non-relativistic entanglement entropy from Ho\v rava gravity}

\author{Stefan Janiszewski}

\affiliation{Department of Physics and Astronomy, University of Victoria, \\ Victoria, BC, V8W 3P6, Canada}

\emailAdd{stefanjj@uvic.ca}

\abstract{We propose an analogue of the Ryu-Takayanagi formula for holographic entanglement entropy applicable to non-relativistic holographic dualities involving Ho\v rava gravity. This is a powerful tool for the duality to have, as topological order quantified by entanglement entropy is a robust notion in condensed matter systems. Our derivation makes use of examining on-shell gravitational actions on conical spacetimes.}

\begin{document}
\maketitle
\flushbottom

\newpage

\section{Introduction}

Entanglement entropy is a universal property of systems described by a Hilbert space \cite{Amico:2007ag}. Given a state $|\Psi\rangle$ in the Hilbert space, and a partition of a complete set of observables into two subsystems, say $A$ and $B$, one can define the reduced density matrix of $A$ by tracing over a basis for $B$:
\begin{eqnarray}
\rho_A\equiv\text{Tr}_B|\Psi\rangle\langle\Psi|.
\label{eq:reddens}
\end{eqnarray}
The entanglement entropy of subsystem $A$, $S_A$, is defined as the von Neumann entropy of this density matrix:
\begin{eqnarray}
S_A\equiv -\text{Tr}_A \hat{\rho}_A \log \hat{\rho}_A,
\label{eq:see}
\end{eqnarray} 
where the trace is now over a basis of states in $A$, and $\hat{\rho}_A\equiv\rho_A/\text{Tr}_A \rho_A$ to properly normalize the reduced density matrix. The entanglement entropy measures the degree to which the state described by $\rho_A$ is mixed, and, as such, it quantifies the entanglement between the two subsystems $A$ and $B$. If the total state is a product state over the subsystems, $|\Psi\rangle =|\Psi_A\rangle|\Psi_B\rangle$, then the entanglement entropy vanishes, otherwise it is greater than zero.

Entanglement entropy is an important observable as it is defined for all quantum systems, regardless of details such as interactions or symmetries\footnote{This does not mean it is easy to calculate. For example, difficulties arise when gauge symmetries are involved \cite{Casini:2013rba}.}. It therefore has a wide range of application from condensed matter systems to quantum field theory. In particular, it has been proposed as an order parameter for phase transitions that lack any traditional local order parameter \cite{Kitaev:2005dm,2006PhRvL..96k0405L}. Situations include quantum critical points, as the entanglement entropy does not vanish at zero temperature, and topological phases, as the entropy is a measure of non-local entanglement that local correlation functions fail to see. An example is the topological order seen in the fractional quantum Hall effect.

\begin{figure}%
\begin{center}\includegraphics[width=.5\columnwidth]{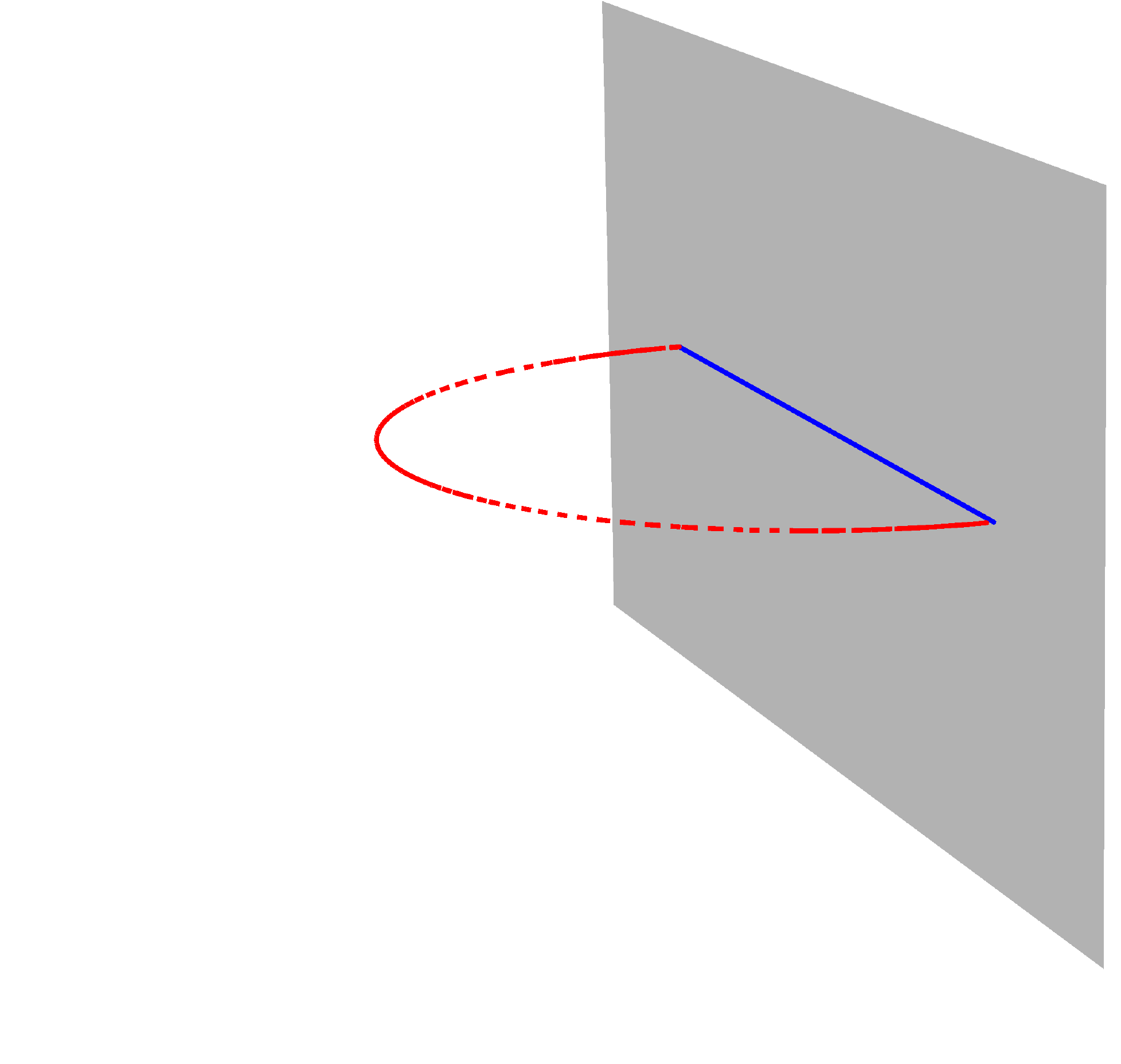}%
\caption{Holographic calculation of entanglement entropy for the spatial region $A$ (dashed, blue) of the boundary spacetime (gray) is given by the area of a bulk surface $\tilde A$ (dashed, red).}%
\label{fig:minsur}%
\end{center}\end{figure}

In many applications, the subsystems $A$ and $B$ are chosen to be complementary spatial subregions of the full system. In most cases, the entanglement entropy then obeys an ``area law'', that is, its leading contribution is proportional to the volume of the boundary of the region $A$, $\partial A$ \cite{Srednicki:1993im}:
\begin{eqnarray}
S_A=\gamma \text{Area}(\partial A)+\cdots,
\label{eq:area}
\end{eqnarray}
where $\gamma$ is a non-universal UV divergent coefficient, and $\cdots$ indicate subleading terms, including finite universal contributions. The form of Eq.~(\ref{eq:area}) bears a striking resemblance to the entropy of a black hole in general relativity (GR) \cite{Bekenstein:1973ur}:
\begin{eqnarray}
S_{BH}=\frac{A_H}{4 G_N},
\label{eq:bhe}
\end{eqnarray}
where $A_H$ is the area of the event horizon, and $G_N$ is Newton's gravitational constant. Indeed, a connection between these notions of entropy is found within gauge/gravity duality \cite{Maldacena:1997re}: for a holographic quantum system at the boundary of its dual spacetime, the entanglement entropy of its spatial subregion $A$ is given by GR as \cite{Ryu:2006bv}:
\begin{eqnarray}
S_A=\frac{\text{Area}(\tilde A)}{4G_N},
\label{eq:rt}
\end{eqnarray} 
where $\tilde A$ is the minimal area surface in the bulk spacetime that shares the boundary of $A$, $\partial \tilde A=\partial A$. See Figure \ref{fig:minsur} for an illustration of the situation. 
Such holographic calculations of the entanglement entropy have been precisely checked against quantum mechanical derivations, when available, and have yielded new predictions otherwise \cite{Nishioka:2009un}.

The current paper aims to propose a formula analogous to Eq.~(\ref{eq:rt}) for use in non-relativistic holography featuring Ho\v rava gravity \cite{Janiszewski:2012nb}. The low energy regime of this duality features a geometric gravitational theory of a spacetime equipped with an additional co-dimension 1 foliation by a global time, which subsequently implies a more restrictive class of diffeomorphism invariance than general relativity. This invariance can capture the generic global symmetries of non-relativistic quantum field theories \cite{2006AnPhy.321..197S,Son:2008ye} (and much of the recent Newton-Cartan structure of \cite{Son:2013rqa,Jensen:2014aia}) which motivated the holographic construction of \cite{Janiszewski:2012nb}. Ho\v rava gravity is reviewed in Section \ref{sec:hgrav}. Many of the systems where entanglement entropy has been proposed as a useful tool are non-relativistic, motivating the understanding of a holographic implementation in Ho\v rava gravity. 

Section \ref{sec:nree} presents the logical underpinning of the conjectured entanglement entropy formula, which follows \cite{Lewkowycz:2013nqa,Dong:2016hjy}. This involves the calculation of the euclidean Ho\v rava gravity action on conical spaces, where we must be especially careful to avoid possible complications in the euclideanization process. This argument leads to the proposal that the entanglement entropy of a subregion $A$ of a holographic non-relativistic quantum field theory is given by its Ho\v rava gravity dual as:
\begin{eqnarray}
S^H_A=\sqrt{1+\beta}\left(1-\frac{\alpha}{1+\beta}\right)\frac{\text{Area}(\tilde A)}{4 G_H}=\frac{s_2}{z}\frac{\text{Area}(\tilde A)}{4 G_H},
\label{eq:areah}
\end{eqnarray}
where $G_H$ is the gravitational constant of Ho\v rava gravity, $\alpha$ and $\beta$ are coupling constants, and $\tilde A$ is the bulk spatial surface of minimal area at the same fixed global time as $A$ (and shares its boundary). The second equality expresses the bulk coupling constants physically in terms of the speed of the spin-2 graviton $s_2=\sqrt{1+\beta}$, and the dynamical critical exponent $z$, $\alpha/(1+\beta)=(z-1)/z$, which controls the anisotropic scaling of time versus space in the non-relativistic field theory \cite{Janiszewski:2014iaa}. 

Section \ref{sec:repae} contains the main justification for the proposal Eq.~(\ref{eq:areah}). The logic takes advantage of the replica trick to calculate entanglement entropy and follows methods used in general relativity \cite{Lewkowycz:2013nqa,Dong:2016hjy}, reviewed in Section \ref{sec:repgr}. The main calculation is of the on-shell gravitational action on various conical spaces. The second piece of justification is discussed in Section \ref{sec:amin} and concerns exactly why $\tilde A$ is a minimal spatial surface. In general relativity the holographic entanglement entropy Eq.~(\ref{eq:rt}) was originally proposed for static spacetimes where $\tilde A$ is taken to be at a constant slice of Killing time. This restriction is necessary on a Lorentzian manifold, as bending a surface into a light-like direction will reduce its area. Later, Eq.~(\ref{eq:rt}) was presented in a covariant form \cite{Hubeny:2007xt}, using the invariant light cone structure of Lorentzian manifolds. For Ho\v rava gravity, no such structure exists, as there is no longer a finite limiting speed. Instead, causality is enforced by the global time foliation structure: signals can only propagate from one leaf in the foliation to another at a later global time coordinate. Our ``covariant'' Ho\v rava proposal is therefore somewhat simpler in this regard: possible $\tilde A$s to minimize over must be at a constant global time in the bulk, fixed by the time of the boundary subregion $A$, and therefore Eq.~(\ref{eq:areah}) trivially generalizes to time dependent states. As in the GR case \cite{Lewkowycz:2013nqa,Dong:2016hjy} $\tilde A$ is shown to be an extremal surface due to the leading equations of motion near the near the tip of a conical space.

\begin{figure}%
\begin{center}\includegraphics[width=.5\columnwidth]{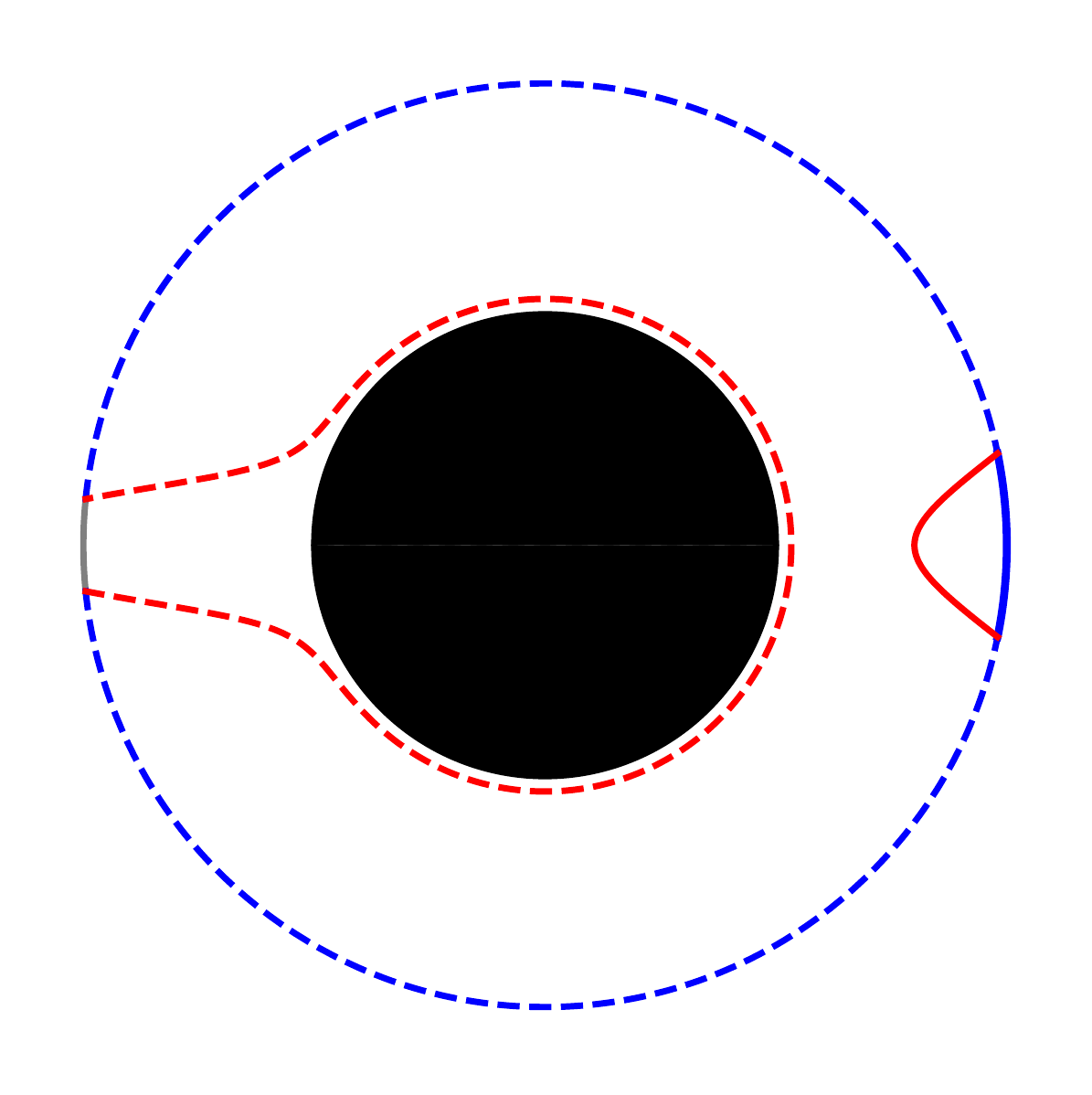}%
\caption{A bulk spacetime with a black hole. The entanglement entropy of the small solid blue boundary region is given by the solid red bulk surface, which is barely influenced by the black hole. For the larger blue boundary region (solid and dashed) its corresponding bulk surface (red, dashed) hugs the black hole horizon, giving a thermal contribution.}%
\label{fig:bh}%
\end{center}\end{figure}

Section \ref{sec:bhee} presents an example of Eq.~(\ref{eq:areah}). This first check uses the fact that for a thermal state $|\Psi_T\rangle$, the entanglement entropy of an infinitely large region tends to the thermal entropy of the system, apart from the UV divergent area term. In the gauge/gravity setting, this translates to the holographic entanglement entropy of an infinitely large boundary region tending to the entropy of the black hole in the bulk spacetime. See Figure \ref{fig:bh} for illustration.
A non-trivial check of the normalization of our proposed holographic entanglement entropy formula is that it reproduces this limiting value for the black hole solution of Ho\v rava gravity of \cite{Janiszewski:2014iaa}.

Section \ref{sec:hee} contains another example making use of Eq.~(\ref{eq:areah}). Parallel to \cite{Bhattacharya:2012mi} we use holographic entanglement entropy to check the universal behavior of the quantum information contained in a field theory region, and its relation to the energy density. We find that the required dimensionality of these quantities, as determined by non-relativistic scaling, is correctly captured in our proposal for holographic entanglement entropy in Ho\v rava gravity.  

\section{Ho\v rava gravity and a black hole}\label{sec:hgrav}

Ho\v rava gravity \cite{Horava:2009uw} is a proposed quantum theory of gravity. Its low energy behavior can be written in terms of the ADM decomposition of a spacetime metric:
\begin{eqnarray}
g_{MN}=\left(\begin{array}{c c}
-N^2+N_KN^K & N_I	\\
N_J & G_{JI}
\end{array}\right),
\label{eq:adm}
\end{eqnarray}
where $G_{IJ}$ is a spatial metric on leaves of a foliation by global time $t$, the lapse $N$ gives the normal distance between leaves, while the shift $N_I$ relates events with the same spatial coordinates, but on different leaves. All spatial indices $I,J,\cdots$ are lowered and raised with the spatial metric $G_{IJ}$ and its inverse. In terms of these fields the extrinsic curvature of the leaves of the foliation is:
\begin{eqnarray}
K_{IJ}\equiv \frac{1}{2N}\left(\partial_t G_{IJ}-\nabla_I N_J-\nabla_J N_I\right),
\label{eq:ext}
\end{eqnarray} while the two derivative Ho\v rava action is:
\begin{eqnarray}
S_H=\kappa\int dtdrd^dx N\sqrt{G}\left(K_{IJ}K^{IJ}-(1+\lambda)K^2+(1+\beta)(R-2\Lambda)+\alpha \frac{\nabla_IN\nabla^I N}{N^2}\right)
\label{eq:act}
\end{eqnarray}
where: $K\equiv K^I_I$; the spatial metric has determinant $G$, Ricci scalar curvature $R$, and associated covariant derivative $\nabla_I$; $\Lambda$ is a cosmological constant; and $\kappa\equiv 1/(16 \pi G_H)$ for $G_H$ the Ho\v rava gravitational constant, while $\alpha$, $\beta$, and $\lambda$ are dimensionless coupling constants. The action Eq.~(\ref{eq:act}) has spatial diffeomorphisms, $x_I\to\tilde x_I(t,x_J)$, and temporal reparametrizations, $t\to \tilde t(t)$, as its gauge symmetries. 

The action Eq.~(\ref{eq:act}) describes the dynamics of spin-2 and spin-0 graviton modes. By examining linear perturbations about the flat background $G_{IJ}=\delta_{IJ}$, $N_I=0$, and $N=1$, these modes are seen to have the speeds squared:
\begin{eqnarray}
s_2^2=1+\beta, \quad s_0^2=\frac{\lambda(1+\beta)\left(d(1+\beta)-(d-1)\alpha\right)}{\alpha\left((d+1)\lambda+d\right)},
\label{eq:speed}
\end{eqnarray}

The full quantum Ho\v rava action includes all higher derivative terms allowed by symmetries that are relevant \cite{Horava:2009uw}. Holographically, working with just the classical low energy action Eq.~(\ref{eq:act}) means that the dual quantum field theory is in a regime with strong coupling and a large number of degrees of freedom.

In four spacetime dimensions, with a cosmological constant\footnote{This is in units of the curvature radius $L$, which we set to 1.} $\Lambda=-3$ and the coupling $\alpha=0$ there is a black hole solution to the theory Eq.~(\ref{eq:act}), given by \cite{Janiszewski:2014iaa}:
\begin{eqnarray}
G_{IJ}=\left(\begin{array}{c c c}
\frac{r_h^6}{r^2(r_h^3-r^3)^2} & 0 & 0 \\
0 & \frac{1}{r^2} & 0\\
0 & 0 & \frac{1}{r^2}
\end{array}\right),\quad
N_I=\left(\frac{\sqrt{1+\beta}r}{r_h^3-r^3},0,0\right),\quad N=\frac{r_h^3-r^3}{r_h^3r}.
\label{eq:solu}
\end{eqnarray}
This solution has an asymptotic boundary as the radial coordinate $r\to 0$, where the corresponding spacetime metric Eq.~(\ref{eq:adm}) is that of Anti-de Sitter. There is a causal ``universal horizon'' at $r=r_h$, from which behind no signals can propagate to the boundary, no matter their speed. See \cite{Janiszewski:2014iaa} for full details regarding interpreting the solution Eq.~(\ref{eq:solu}) as a black hole of Ho\v rava gravity. Its energy density, temperature, and entropy density are given by:
\begin{eqnarray}
\label{eq:therm}
\epsilon_{BH}=\frac{1+\beta}{4\pi G_H r_h^3},\quad T_{BH}=\frac{3\sqrt{1+\beta}}{2\pi r_h},\quad s_{BH}=\frac{\sqrt{1+\beta}}{4 G_H r_h^2},
\end{eqnarray} 
respectively.

It will prove useful to recast Ho\v rava gravity in a fully spacetime covariant form in order to use the arguments of \cite{Lewkowycz:2013nqa,Dong:2016hjy} concerning holographic entanglement entropy. This can be accomplished by its relation to Einstein-aether theory \cite{Jacobson:2000xp,Barausse:2011pu}. Consider the action of a spacetime metric $g_{MN}$ and unit time-like ``aether'' vector $u_M$:
\begin{eqnarray}
S_{AE}=\frac{1}{16\pi G_{AE}}\int dtdrd^dx\sqrt{-g}\bigg(&\tilde R&-2 \Lambda-c_2\left(\tilde\nabla_Mu^M\right)^2 \nn
&-&c_3\tilde\nabla_Mu^N\tilde\nabla_Nu^M+c_4u^M\tilde\nabla_Mu^Nu^P\tilde\nabla_Pu_N\bigg),
\label{eq:aeth}
\end{eqnarray}
where $g_{MN}$ has determinant $g$, Ricci scalar $\tilde R$, and associated covariant derivative $\tilde\nabla _M$; $G_{AE}$ is a gravitational constant, and $c_i$ are dimensionless coupling constants\footnote{The $c_1$ coupling constant of Einstein-aether theory has been set to $0$ as $u_{[N}\tilde\nabla_P u_{Q]}=0$ since the aether vector will be assumed to be hypersurface orthogonal for application to Ho\v rava gravity \cite{Barausse:2011pu}.}. The action Eq.~(\ref{eq:aeth}) has the full spacetime diffeomorphism invariance of general relativity. To relate to Ho\v rava gravity, one demands that $u_M$ is hypersurface orthogonal, and then partially fixes the coordinate invariance by performing a temporal diffeomorphism so that $u_M$ has only a time component. Then, decomposing the spacetime metric $g_{MN}$ into the ADM fields of Eq.~(\ref{eq:adm}), one has $u_M=-N\delta^{\ t}_M$, and the Einstein-aether action Eq.~(\ref{eq:aeth}) becomes (up to total derivatives) the Ho\v rava action Eq.~(\ref{eq:act}) once the constants are identified as:
\begin{eqnarray}
\frac{G_H}{G_{AE}}=1+\beta=\frac{1}{1-c_3},\quad 1+\lambda=\frac{1+c_2}{1-c_3},\quad \alpha=\frac{c_4}{1-c_3}.
\label{eq:const}
\end{eqnarray} 
The black hole solution Eq.~(\ref{eq:solu}) is therefore a solution to Einstein-aether theory with $\Lambda=-3$ and $c_4=0$.

A property of the Einstein-aether formalism that will prove useful is that the action Eq.~(\ref{eq:aeth}) has a field redefinition invariance \cite{Foster:2005ec}. For the redefined metric and aether vector:
\begin{eqnarray}
\hat g_{MN}\equiv g_{MN}-(\sigma-1)u_Mu_N, \quad \hat u^M\equiv u^M/\sqrt{\sigma},
\label{eq:redef}
\end{eqnarray}
with $\sigma>0$, the action Eq.~(\ref{eq:aeth}) retains its form in terms of these hatted fields, but with $\hat G_{AE}=\sqrt{\sigma}G_{AE}$ and new coupling constants $\hat c_i$, given explicitly in terms of $c_i$ and $\sigma$ in \cite{Foster:2005ec}\footnote{A generic transformation will generate a $\hat c_1$ term, but this can again be set to zero as $\hat u_M$ is still hypersurface orthogonal.}. In particular, for $\sigma$ equal to the spin-2 graviton speed squared, $s^2_2=1+\beta=1/(1-c_3)$, $\hat c_3=0$ and the redefined metric $\hat g_{MN}$ is an effective metric such that the sound horizon for the spin-2 modes is now a Killing horizon. For the Ho\v rava action Eq.~(\ref{eq:act}) this is equivalent to the invariance: $N\to N/\sqrt{1+\beta}$, $\alpha\to (1+\beta)\alpha$, and $G_H\to G_H\sqrt{1+\beta}$.

These redefinitions are useful as they help clarify some subtleties concerning units. Recall that the graviton speeds Eq.~(\ref{eq:speed}) are derived by examining fluctuations about the Minkowski vacuum $\eta_{MN}$. Therefore the dimensionless speed $s_2=\sqrt{1+\beta}$ is being expressed as a multiple of the unit null speed of the Minkowski metric, which has no special meaning in Ho\v rava gravity, it simply converts units of time into units of space. We can physically simplify the situation by choosing time units such that the spin-2 graviton has speed 1, this is equivalent to the field redefinition of the previous paragraph. We can always do this, as long as there is no additional field content in the theory, which would fix a preferred $\hat g_{MN}$ by its coupling. Therefore, from now on our action is Eq.~(\ref{eq:aeth}) with $c_3=0$ and $G_{AE}=G_H/\sqrt{1+\beta}$.
 
\section{Holographic non-relativistic entanglement entropy}\label{sec:nree}

The following sections build up justification for the proposed Eq.~(\ref{eq:areah}) for holographic non-relativistic entanglement entropy. Section \ref{sec:repgr} may be skipped by those with experience deriving Eq.~(\ref{eq:rt}) in traditional holography. 

\subsection{Replica trick in general relativity}
\label{sec:repgr}

\begin{figure}
\centering
\subfloat[][]{
  \includegraphics[width=.4\textwidth]{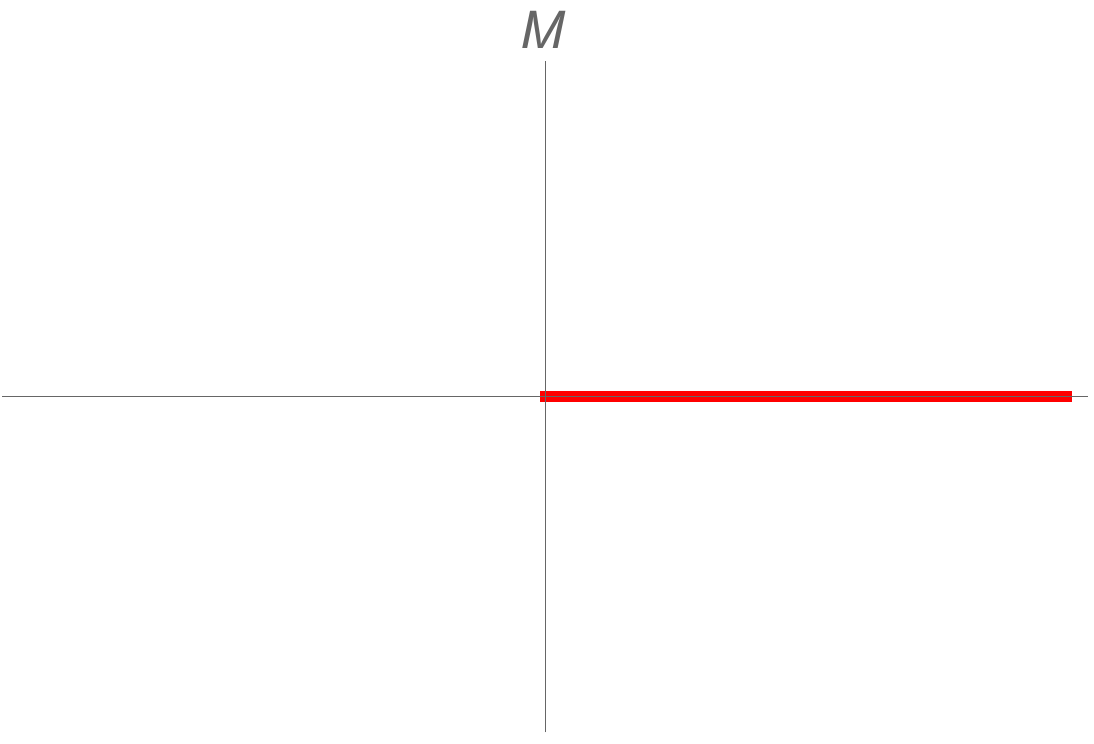}}\\
\subfloat[][]{
  \includegraphics[width=.3\textwidth]{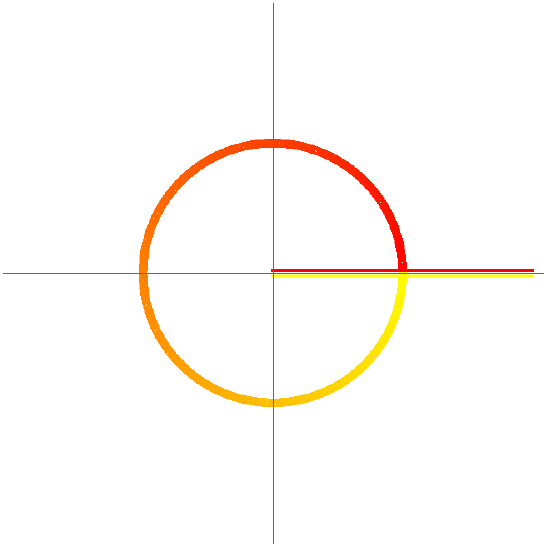}}
\subfloat[][]{
  \includegraphics[width=.3\textwidth]{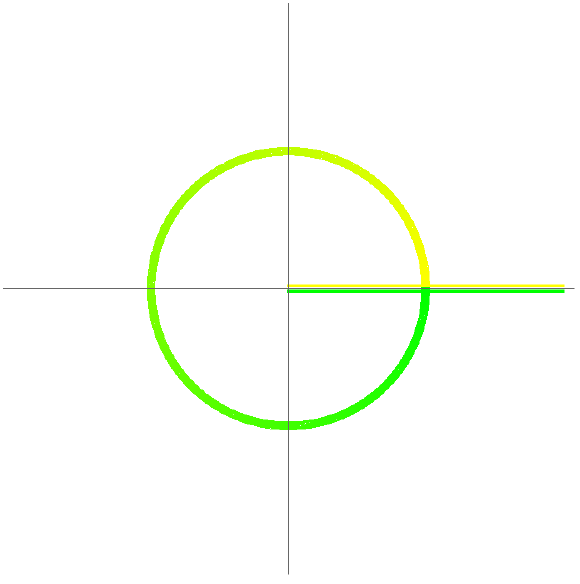}}
\subfloat[][]{
  \includegraphics[width=.3\textwidth]{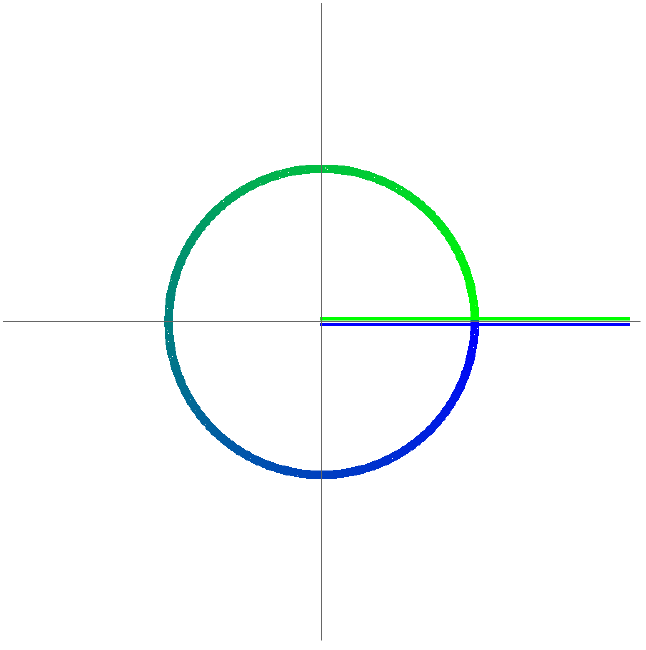}}
\caption{(a) A spacetime $\mathcal M$ with the spatial region $A$, in red. (b)-(d) The replica spacetime $\mathcal M_n$ is created from $n$ copies of $\mathcal M$ by cutting along $A$ and gluing cyclically. As we go counterclockwise from above the cut to below it (red to yellow on (b)) we emerge on the next sheet (the yellow of (c)). This process is repeated $n$ times, while identifying the bottom of the final cut (blue of (d)) with the top of the first (red of (b)) computes the trace in the partition function.}
\label{fig:rep}
\end{figure}

Evidence for the formula Eq.~(\ref{eq:rt}) in general relativity was presented in \cite{Fursaev:2006ih,Lewkowycz:2013nqa,Dong:2016hjy} and takes advantage of what is known as the ``replica trick''. This calculational tool is often used in field theoretic derivations of the entanglement entropy and takes advantage of the identity involving Eq.~(\ref{eq:see}):
\begin{eqnarray}
S_A=-\text{Tr}_A \hat{\rho}_A\log\hat{\rho}_A=-n\frac{\partial}{\partial n}\left(\log \text{Tr}_A\rho_A^n-n\log \text{Tr}_A\rho_A\right)\big|_{n=1}.
\label{eq:rep}
\end{eqnarray} 
In order to calculate the density matrix product, the replica trick uses the fact that the path integral calculation of $\text{Tr}_A\rho_A^n$ is formally equivalent to the euclidean partition function on an $n$-sheeted Riemann surface $\mathcal M_n$ made by gluing together copies of the original manifold $\mathcal M$ cut along the surface $A$ as in Figure \ref{fig:rep}. See, for example, \cite{Nishioka:2009un} for more details.

\begin{figure}
\centering
\subfloat{
  \includegraphics[width=.4\textwidth]{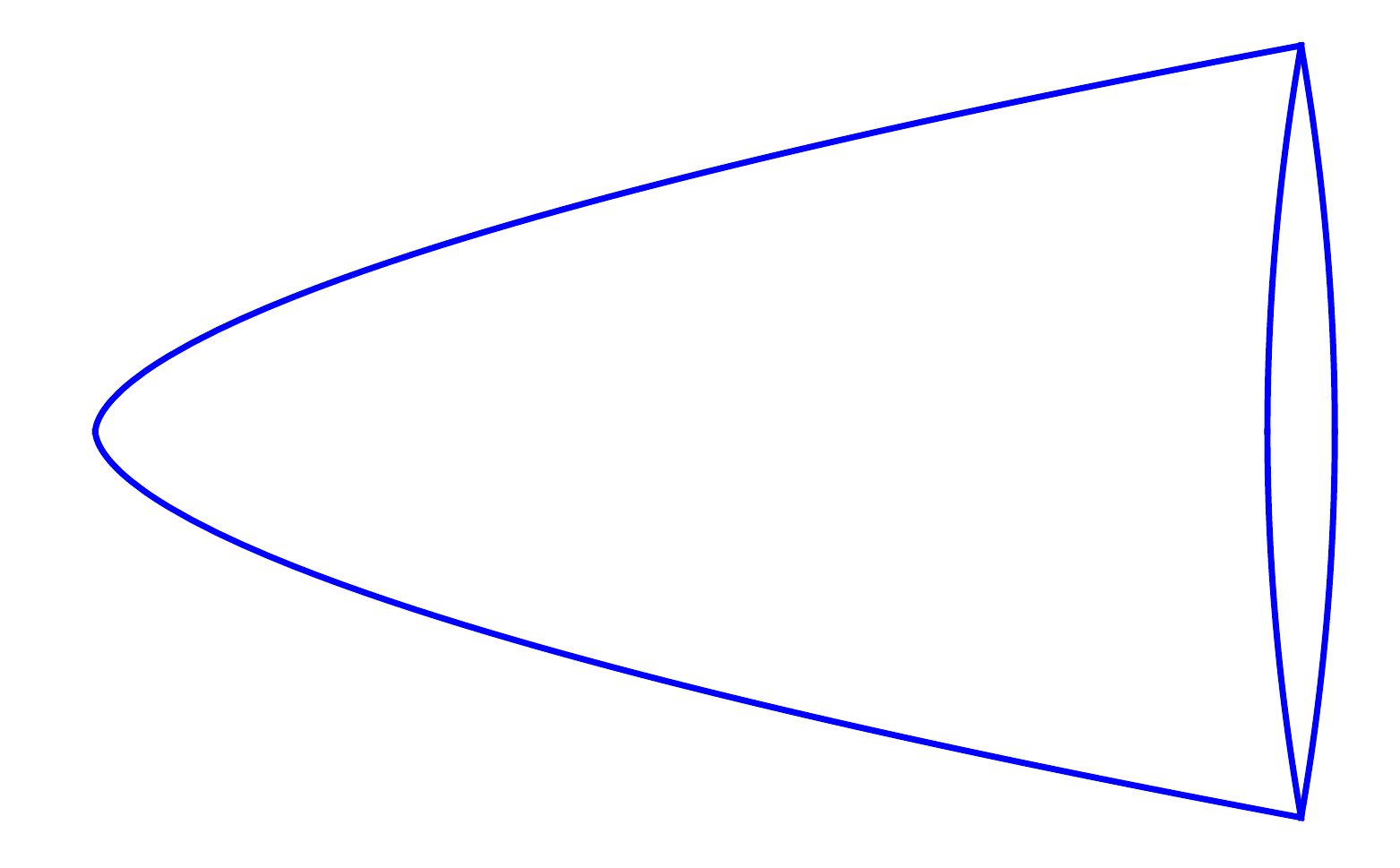}}
\subfloat{
  \includegraphics[width=.4\textwidth]{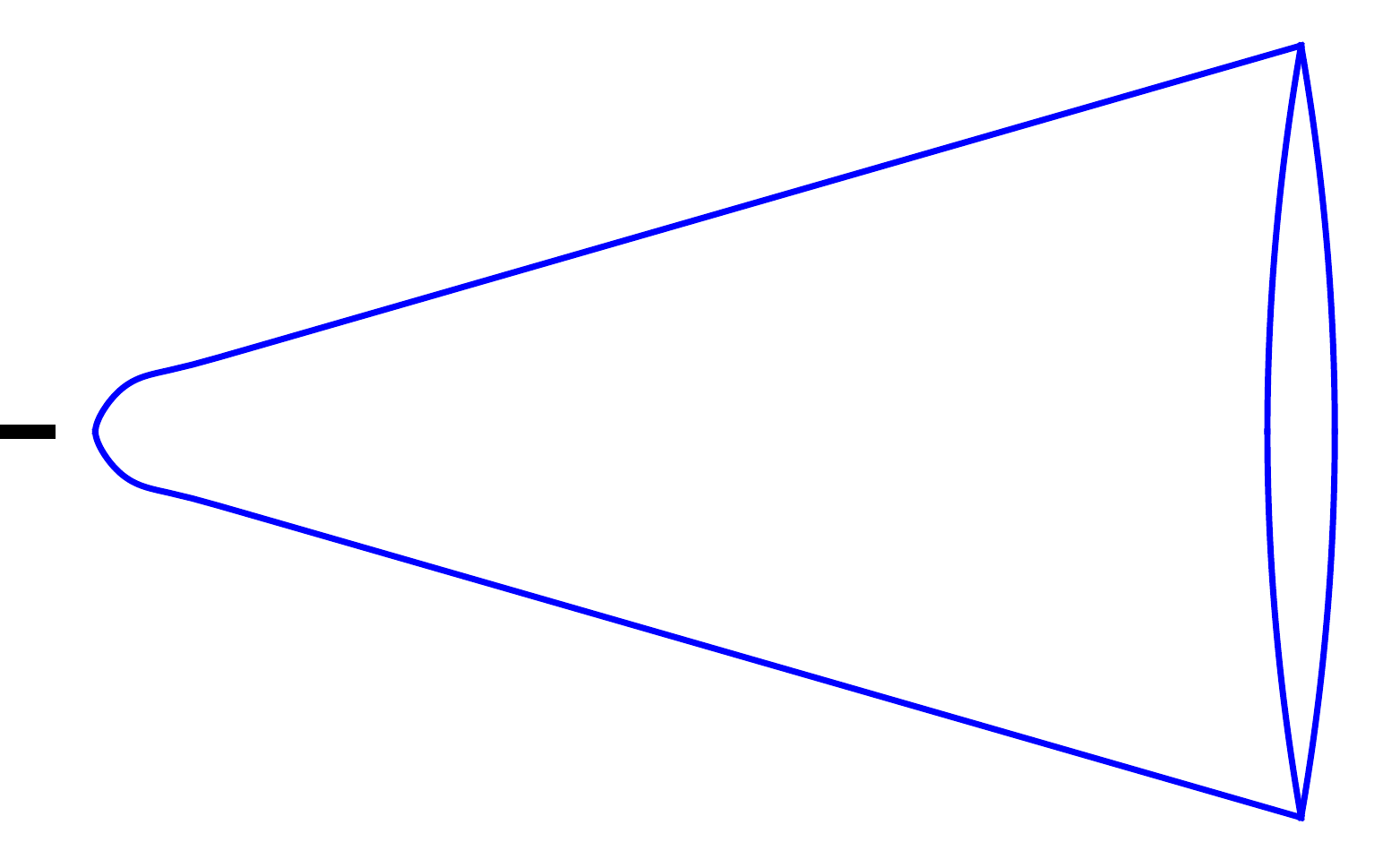}}\\
\subfloat{
  \includegraphics[width=.4\textwidth]{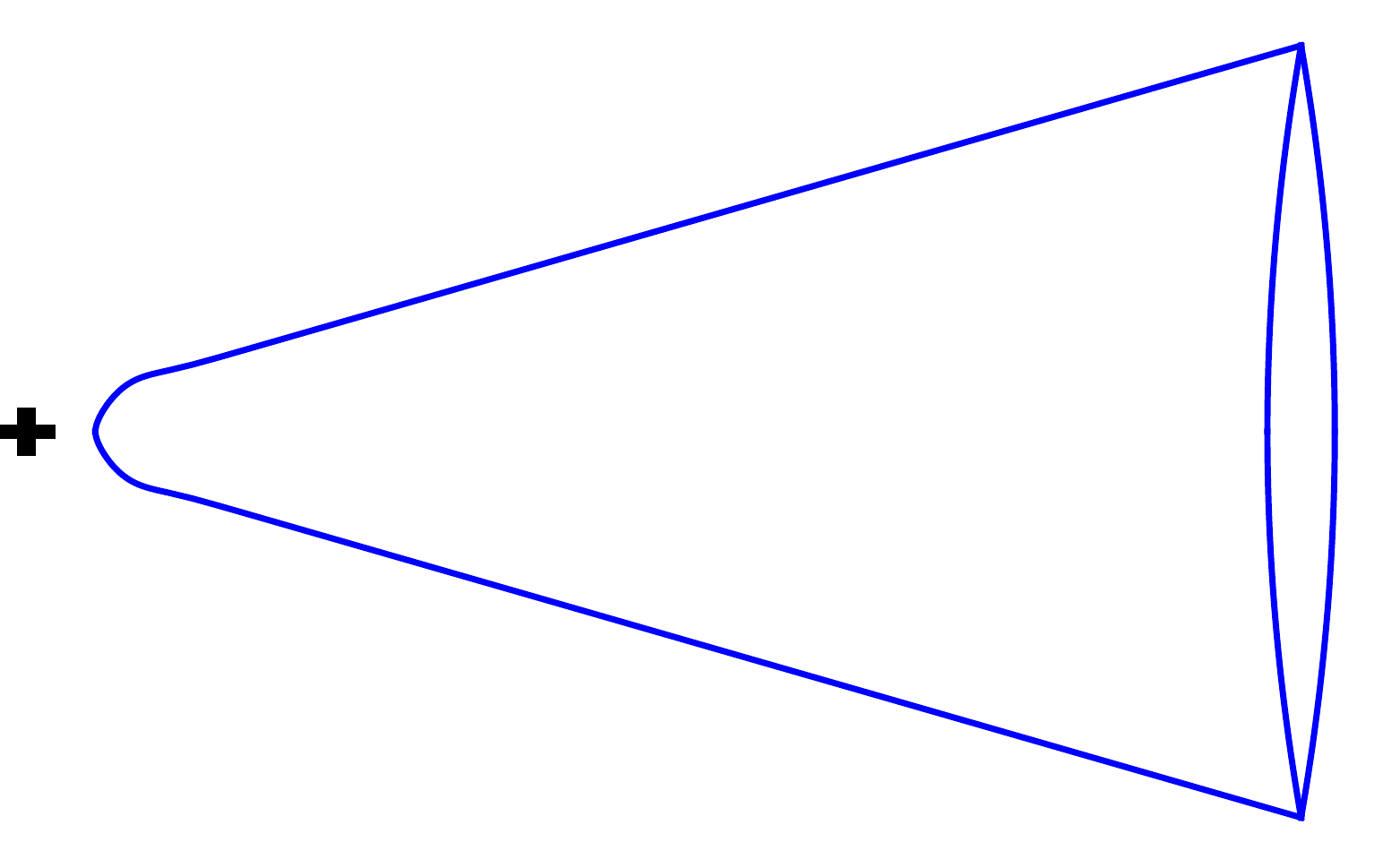}}
\subfloat{
  \includegraphics[width=.4\textwidth]{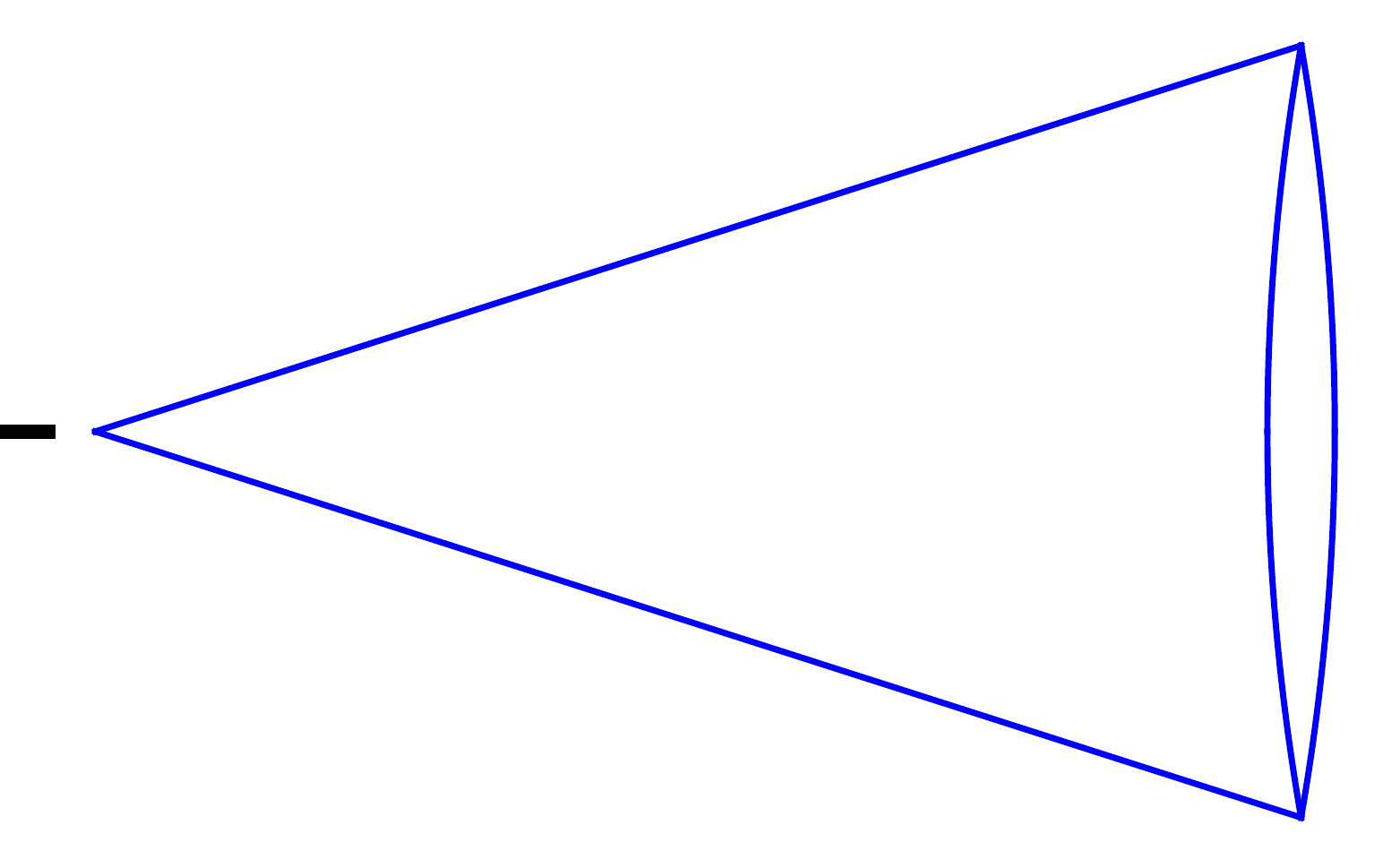}}
\caption{The first term is the correct smooth solution for replica spacetime $n$, the last term is the $n=1$ solution, but with time extent $2\pi n$, and therefore has a conical singularity. These two terms can compute the partition functions of Eq.~(\ref{eq:rep}). The middle two terms are regulated versions of the cone. For $n\approx 1$, the two terms of the first line cancel, as the regulated cone is a first order variation away from the true $n$ solution. The two terms of the bottom line only give a contribution from the tip of the regulated cone, and are the terms used to compute Eq.~(\ref{eq:conereg}).}
\label{fig:cone}
\end{figure}

To holographically calculate the entanglement entropy, this logic is extended to the gravitational bulk \cite{Fursaev:2006ih,Lewkowycz:2013nqa}. To calculate the $n$-sheeted partition function, one can relate it to the on-shell euclidean gravitational action $I_{GR}^E$ on a bulk spacetime $\tilde{\mathcal{M}}_n$ whose asymptotic boundary is conformal to $\mathcal M_n$. Although the construction of $\tilde{\mathcal{M}}_n$ for general $n$ remains to be fully understood, near the limit $n\approx1$ required for the entanglement entropy calculation, the formula Eq.~(\ref{eq:rep}) can be expressed as \cite{Lewkowycz:2013nqa}:
\begin{eqnarray}
S_A=-n\frac{\partial}{\partial n}\left(-I_{GR}^E(n,\text{reg})+I_{GR}^E(n,\text{cone})\right)\big|_{n=1},
\label{eq:conereg}
\end{eqnarray}
where: $I^E_{GR}(n,\text{cone})$ is the euclidean general relativity action evaluated on the solution $\tilde{\mathcal{M}}_{n=1}$, except that the euclidean time is integrated to $2\pi n$, and therefore the space is a cone with surplus angle $2\pi(n-1)$; $I^E_{GR}(n,\text{reg})$ is the euclidean general relativity action evaluated on a regularized version of this cone. Figure \ref{fig:cone} illustrates the equivalence of Eq.~(\ref{eq:rep}) and Eq.~(\ref{eq:conereg}) for $n\approx1$, see \cite{Lewkowycz:2013nqa} for further details.

In comparing the euclidean GR actions of the cone and the regularized cone, the only surviving contribution comes from the tip of the regularized cone due to the fact that on such a space the curvature integral gives:
\begin{eqnarray}
\int dr \sqrt{g}R=2(1-n)+\mathcal{O}((n-1)^2,a),
\label{eq:intregcone}
\end{eqnarray} 
where $r$ is a coordinate away from the cone tip, $\mathcal{O}((n-1)^2)$ indicates higher order terms for $n\approx 1$, and $\mathcal{O}(a)$ indicates terms that vanish as the regularization of the cone is taken away. Finally, we see that Eq.~(\ref{eq:conereg}) becomes Eq.~(\ref{eq:rt}) for the holographic entanglement entropy:
\begin{eqnarray}
S_A=-\frac{n}{16\pi G_N}\frac{\partial}{\partial n}\left(\int_0^{2\pi n} d\tau \int dx^d2(1-n)\right)\big|_{n=1}=\frac{\text{Area}(\tilde A)}{4G_N},
\label{eq:heeproof}
\end{eqnarray}
where $\tilde A$ is the bulk surface tangential to the tip of the regularized cone. This can also be calculated from the boundary contributions from a conical spacetime \cite{Lewkowycz:2013nqa}, both of which we do for Ho\v rava gravity in the next section. The fact that the area of $\tilde A$ is minimal is due to the leading order Einstein's equations \cite{Lewkowycz:2013nqa}.

As presented, these arguments are most rigorous for static spacetimes where there is no impediment to euclideanization and the path integral calculation of the density matrix is a sensible procedure. Covariantization of these arguments to general time dependent spacetimes was carried out for GR in \cite{Dong:2016hjy}. In the following sections we will see that for Ho\v rava gravity the general form of holographic entanglement entropy is much simpler to justify. 

\subsection{Replica trick in Einstein-aether}
\label{sec:repae}

For Einstein-aether gravity a similar argument as presented in the last section can be attempted. Now that the theory content contains a vector we must understand its transformation in the euclideanization process. This can be determined by expressing the aether vector in terms of a scalar field, of which it is the gradient. In this context the scalar $\phi$ is called the khronon \cite{Germani:2009yt,Blas:2010hb}, as its level sets determine the foliation by a global time. Taking into account normalization, the aether vector can be written as:
\begin{eqnarray}
u_M\equiv\frac{-\partial_M\phi}{\sqrt{-g^{NP}\partial_N\phi\partial_P\phi}}.
\label{eq:khron}
\end{eqnarray}
Therefore, for standard euclideanization with $t=-i \tau$ we see that we require $u_t=-u^E_\tau$ and $u_I=i u_I^E$, where the $E$ subscript signifies euclidean objects. Since the khronon is identified as a time coordinate we have defined $\phi\equiv -i \phi_E$\footnote{This somewhat odd transformation does not change the result: $\phi\equiv\phi_E$ gives the same euclidean action.}.

Given this prescription we can euclideanize the Einstein-aether action Eq.~(\ref{eq:aeth}) and use its value on conical spacetimes to calculate the entanglement entropy from Eq.~(\ref{eq:conereg}). The euclidean action has the signs of the $c_2$ and $c_3$ terms flipped, but all factors remain real. Near the tip of a regularized cone with euclidean time extent $2\pi n$ the metric is:
\begin{eqnarray}
ds_E^2=R^2d\tau^2+n^2dR^2+d\vec{x}^2,
\label{eq:rindn}
\end{eqnarray}
where $R$ is the distance from the origin. The Einstein-aether equations of motion determine the aether vector to behave as:
\begin{eqnarray}
u^E_M=(\sqrt{R^2+(b/n)^2},i b/R,\vec{0}),
\label{eq:coneae}
\end{eqnarray}
near the cone tip, where $b$ is a constant. The cone space of Eq.~(\ref{eq:conereg}) is these fields with $n=1$, but euclidean time extent still given by $2\pi n$, while the regularized cone looks like Eq.~(\ref{eq:rindn}) near its tip but becomes the $n=1$ cone for larger $R$. Recalling the discussion of choice of time units at the end of Section \ref{sec:hgrav}, we see that the Einstein-aether theory gives the entanglement entropy by evaluating the actions of Eq.~(\ref{eq:conereg}) as:
\begin{eqnarray}
S^H_A=\frac{(1-c_4)}{4 G_{AE}}\text{Area}(\tilde A)=\frac{\sqrt{1+\beta}}{4 G_{H}}\left(1-\frac{\alpha}{1+\beta}\right)\text{Area}(\tilde A),
\label{eq:heedev}
\end{eqnarray}
where $\tilde A$ is the surface transverse to the cone tip.

We can confirm this value by repeating the calculation of the entanglement entropy $S^H_A$ via the method of boundary terms from apparent conical singularities \cite{Lewkowycz:2013nqa}. This method relies on the fact that due to the replica symmetry $I^E(n,2\pi n)=n I^E(n,2\pi)$ where $I^E(n,2\pi)$ is the euclidean action evaluated with the smooth replica solution fields labeled by $n$, but with time only integrated over a range of $2\pi$; it therefore appears to be the action of a spacetime with a conical singularity. The replica formula for entanglement entropy Eq.~(\ref{eq:rep}) then becomes \cite{Lewkowycz:2013nqa}:
\begin{eqnarray}
S_A=n^2\frac{\partial}{\partial n}I^E(n,2\pi)|_{n=1}.
\label{eq:eecone}
\end{eqnarray} 
This derivative with respect to $n$ can be considered a variation of the action. Since $n=1$ is a solution, the bulk term of the variation, proportional to the equations of motion, will vanish in the $n=1$ limit. Therefore, there will only be a boundary term contribution at the cone tip, as changing $n$ changes the boundary conditions for the fields at the conical singularity. For Einstein-aether theory this yields \cite{jishnu}:
\begin{eqnarray}
S^H_A=-\frac{1}{8 G_{AE}}\int_{R=0}d^d x\sqrt{g}\bigg(g^{MN}n^P\nabla_P \frac{\partial}{\partial n}g_{MN}-n^M\nabla^N\frac{\partial}{\partial n}g_{MN}+\nonumber\\
\frac{\partial}{\partial n}g_{MN}\left(n_PX^{PMN}-2n_PY^{PM}u^N\right)+2n_MY^{MN}\frac{\partial}{\partial n}u_{N}\bigg),
\label{eq:btcone}
\end{eqnarray}
where $n^M$ is the unit normal to hypersurfaces of constant $R$, and we have defined the aether dependent contributions:
\begin{eqnarray}
Z_{MPN Q}&\equiv& c_2 g_{MN}g_{PQ}+c_3 g_{MQ}g_{PN}-c_4 u_Mu_Pg_{NQ},\\
Y^M_{\hphantom{M} N} &\equiv& Z^{MP}_{\hphantom{MP} N Q}\nabla_Pu^Q,\\
X^P_{\hphantom{P} MN}&\equiv& Y^P_{\hphantom{P} (M}u_{N)}-u_{(M}Y_{N)}^{\hphantom{N}\hphantom{)}P}+u^PY_{(MN)}.
\label{eq:constae}
\end{eqnarray}
Near the origin $R=0$ of the replica spacetime the metric and aether vector behave as Eq.~(\ref{eq:rindn}) and Eq.~(\ref{eq:coneae}), respectively. Plugging these fields into Eq.~(\ref{eq:btcone}) and taking the needed limits gives the same value for $S^H_A$ as the regulated cone method, Eq.~(\ref{eq:heedev}).

As in Section \ref{sec:repgr} these arguments are most rigorous when the spacetime is static. Einstein-aether theory has the further complication that the vector aether field is present, which allows the time direction determined by $u_M$ to be different than that determined by the Killing time. Examining the form of $u_M$ near the cone tip Eq.~(\ref{eq:coneae}), we see that these two notions of time are only aligned when $b=0$, and our calculation should only be trusted for backgrounds in that regime. Luckily, our final answer Eq.~(\ref{eq:heedev}) is independent of $b$, so there is some hope that it extends to general backgrounds, a sentiment borne out next section. 

\subsection{The minimal nature of $\tilde A$}
\label{sec:amin}

The calculations of the previous section give the value of the the non-relativistic holographic entanglement entropy $S^H_A$ in terms of the area of the surface $\tilde A$, transverse to the tip of the cone. Those procedures do not tell us the full nature of this surface. By construction via the replica trick, $\tilde A$ is anchored at the holographic boundary at the edges of $A$, the field theory surface of which we are calculating the entanglement entropy. How it then extends into the bulk spacetime is undetermined by the previous discussion.

In \cite{Dong:2016hjy} it is argued, for Lorentzian general relativity, that for a given Cauchy slice of the boundary containing $A$, $\tilde A$ must be in its Wheeler-DeWitt patch, as $\tilde A$ should be on a bulk Cauchy slice that ends on the boundary one. Furthermore, in order to respect the causality of the boundary, $\tilde A$ must not be in bulk causal contact with either the domain of dependence of $A$, $D[A]$, or that of its complement, $D[A^c]$. This narrows down the location of the cone tip, $\tilde A$, but it is only fixed once the next order equations of motion are imposed \cite{Lewkowycz:2013nqa,Dong:2016hjy}, where it is seen to be extremal, agreeing with the covariant constructions of \cite{Hubeny:2007xt}.

We can follow this logic in application to Ho\v rava gravity. The first interesting difference is that for a generic non-relativistic field theory the domain of dependence of a spatial region $A$ is just the region itself, $D[A]=A$, since signals can travel with arbitrarily high speed. For the bulk to respect the causality of the boundary, if $A$ (and $A^c$) are at the global time $t_*$, then the bulk surface $\tilde A$ must also be at the same fixed global time, otherwise it would be in causal contact with e.g.~$D[A]$ via some fast bulk signal. This fact that the leaf of the boundary foliation at $t_*$ can be uniquely extended into a bulk global time slice is much simpler than in GR, and due to the reduced symmetries of Ho\v rava covariance. In fact, it seems nonsensical to define a ``surface'' in Ho\v rava gravity to be anything but at a fixed global time. 

This narrows down $\tilde A$ to be a spatial surface at a fixed global time, to further classify it we must apply the equations of motion to next order. Following the real-time constructions of \cite{Dong:2016hjy} the first order expansion of the metric away from $\tilde A$ is, in Rindler coordinates:
\begin{eqnarray}
ds^2=-R^2dT^2+n^2dR^2+(h_{ij}+2R^n \text{Cosh}(T) K^x_{ij}+2R^n\text{Sinh}(T)K^{t_m}_{ij})dx^idx^j,
\label{eq:1stord}
\end{eqnarray} 
where $h_{ij}$ is the metric on $\tilde A$ at $R=0$, and $K^x_{ij}$ and $K^{t_m}_{ij}$ are the extrinsic curvatures of $\tilde A$ in the Minkowski directions $(t_m,x)$. In these coordinates, the aether vector near $R=0$ is:
\begin{eqnarray}
u_M=(-\sqrt{R^2+(b/n)^2},b/R,\vec{0}),
\label{eq:aerind}
\end{eqnarray}
for $b$ a constant. 

\begin{figure}%
\begin{center}\includegraphics[width=.2\textwidth]{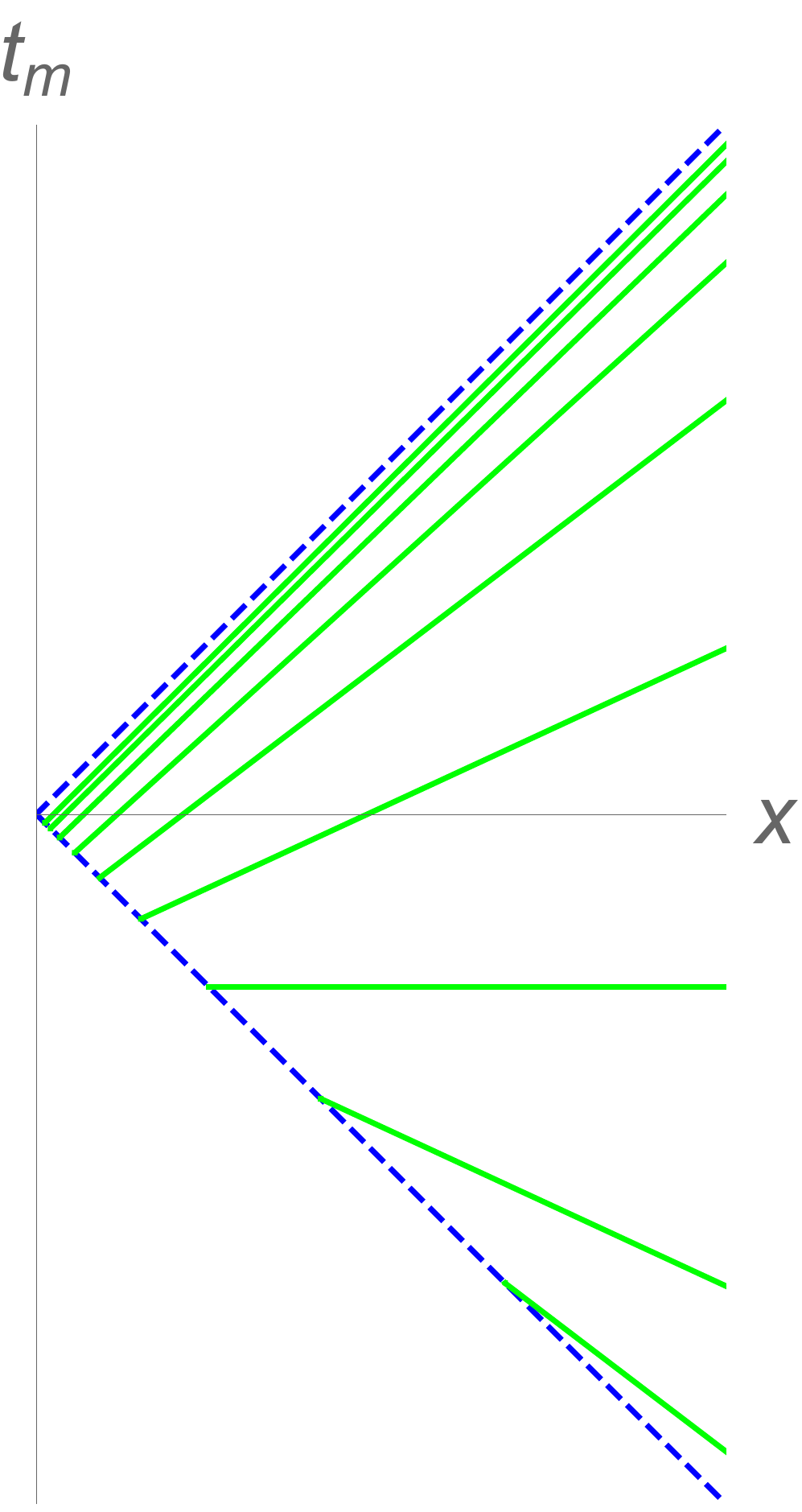}%
\caption{The foliation of the Rindler wedge, bounded by the dashed blue horizons, by the global time, solid green lines.}%
\label{fig:rindt}%
\end{center}\end{figure}

We can immediately see an issue when applying this situation to Ho\v rava gravity: the Rindler time $T$ is not a global time of the theory, as $u_M$ has no spatial components when written in a global time. This means that if we examine the limit $R\to 0$ while holding $T$ constant we will be jumping to different leaves of the global time foliation, while by above we expect $\tilde A$ to reside on one leaf. To remedy this situation we can perform a temporal diffeomorphism to a global time $t\equiv T+h(R)$, where $h(R)$ is determined to eliminate $u_R$. Near $R=0$ this is solved by $h(R)=-n \log R$. Figure \ref{fig:rindt} plots the foliation of the Rindler wedge by this global time, from which it is easy to see that the origin $\tilde A$ is at $t\to\infty$.
We are now in a position to examine the equations of motion for the replica geometry Eq.~(\ref{eq:1stord}) for $n\approx 1$ and $R\to 0$, $t\to\infty$\footnote{Being fully covariant the equations of motion do not non-trivially depend on our coordinate system, we simply find it easier to interpret the following constraint on the extrinsic curvatures in terms of the global time $t$, rather than Rindler time $T$ or Minkowski time $t_m$.}. For Einstein-aether theory we see that:
\begin{eqnarray}
\text{EOMs}\propto h^{ij}(K^x_{ij}+K^{t_m}_{ij})\frac{e^{t}}{R^2}(n-1),
\label{eq:mintr}
\end{eqnarray}
Therefore, as we go away from $\tilde A$ we require that the trace of the $t_m+x$ combination of extrinsic curvatures vanishes. Examining Figure (\ref{fig:rindt}) it is apparent that near $\tilde A$ the $t_m+x$ axis is a slice of constant global time $t$. The requirement $ h^{ij}(K^x_{ij}+K^{t_m}_{ij})=0$ therefore implies that in the leaf of constant global time $\tilde A$ is a minimal surface, justifying the claims surrounding the formula for non-relativistic holographic entanglement entropy Eq.~(\ref{eq:areah}). 

\section{Entanglement entropy of a thermal state}
\label{sec:bhee}

As an example, and check, we will now calculate the entanglement entropy for an infinitely long strip in the non-relativistic field theory dual to the  Ho\v rava gravity black hole Eq.~(\ref{eq:solu}). This boundary strip $A$ will be at a constant time $t_*$, cover $-l\le x\le l$ and have infinite extent in the $y$-direction. To use the proposed holographic formula Eq.~(\ref{eq:heedev}) we need to calculate the area of a bulk spatial surface that shares the boundaries of $A$ at $x=\pm l$. This two-surface $\tilde A$ is on the bulk leaf of global time labeled by $t_*$ and has a spatial profile given by the embedding:
\begin{eqnarray}
X^I=(r,g(r),y),
\label{eq:emb}
\end{eqnarray}
where $r$ and $y$ are used as parametrizing coordinates and $g(r=0)=\pm l$. The area functional of $\tilde A$ is given by:
\begin{eqnarray}
\text{Area}(\tilde A)=\int dr dy \sqrt{h},
\label{eq:areafunc}
\end{eqnarray}
where $h$ is the determinant of the induced metric on $\tilde A$: $h_{ab}\equiv \partial_a X^I\partial_b X^J G_{IJ}$, where $a$, $b$ run over the parametrizing coordinates $r$ and $y$.

\begin{figure}%
\begin{center}\includegraphics[width=.618\columnwidth]{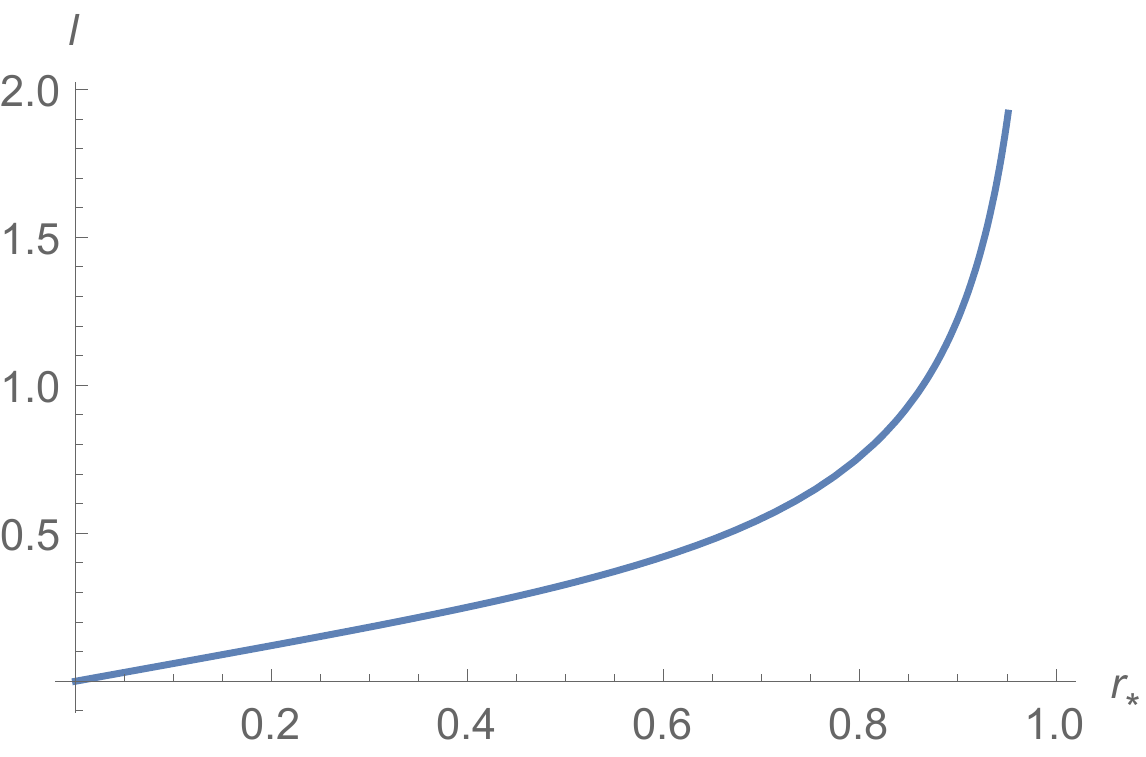}%
\caption{The boundary value $g(r=0)\equiv l$, which is the half-width of the strip $A$, vs the parameter $r_*$. All units are in terms of the universal horizon radius $r_h$. It is evident that $l\to \infty$ as $r_*\to 1$.}%
\label{fig:l}%
\end{center}\end{figure}

\begin{figure}%
\begin{center}
\includegraphics[width=.618\columnwidth]{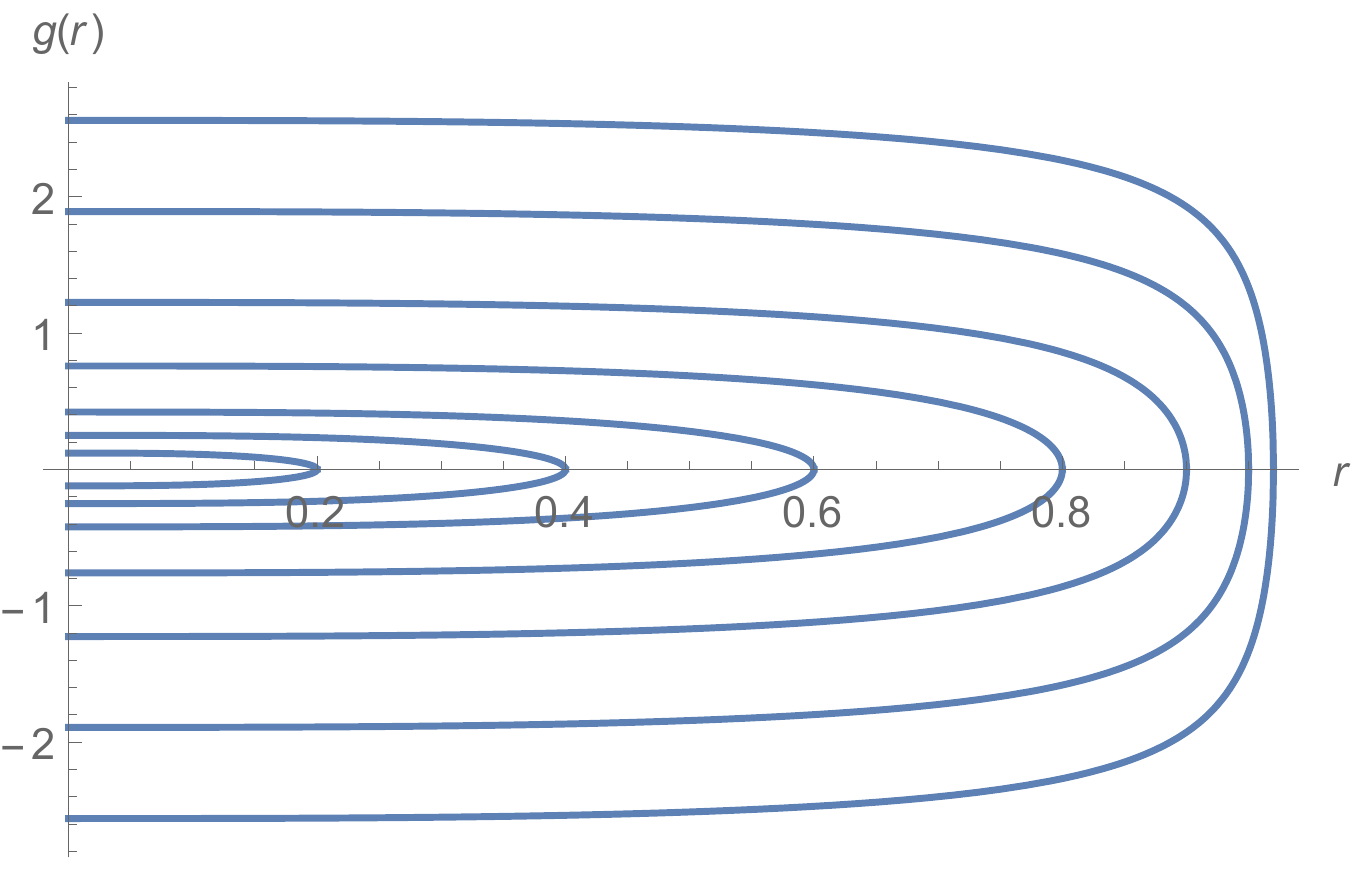}%
\caption{Various bulk surfaces $\tilde A$ as determined by the embedding function $x=g(r)$. All units are in terms of the universal horizon radius $r_h$. Where the different surfaces cap off is given by their value of $r_*$.}%
\label{fig:surf}%
\end{center}\end{figure}

For the above embedding functions Eq.~(\ref{eq:emb}), variation of the area functional Eq.~(\ref{eq:areafunc}) gives the equation of motion for $g(r)$:
\begin{eqnarray}
g'(r)=\frac{\pm r^2}{(1-r^3)\sqrt{r_*^4-r^4}},
\label{eq:eom}
\end{eqnarray}
where $r_*$ is a constant, and we have rescaled coordinates $r\to r_h r$, $y\to r_h y$, etc.~in order to eliminate $r_h$ from the equation. The universal horizon is now at $r=1$. This equation can be analytically integrated to give $g(r)$ in terms of elliptic-$\pi$ integrals and logarithms, but the full form is long and unenlightening. General properties of $g(r)$ can be understood by looking at Eq.~(\ref{eq:eom}): the surface $\tilde A$ starts off normal to the boundary at $r=0$ and then dips into the bulk, but since $g'$ diverges at $r_*$ the surface cannot penetrate that radius, and caps off instead. While $g'$ also diverges at the universal horizon, $r=1$, we will only be concerned with the class of surfaces with $r_*\le 1$, as they can be seen to be the minimal area surface for any size boundary strip. Implementing the boundary condition $g(r=0)=\pm l$ establishes a relation between $r_*$ and $l$, which is plotted in Figure (\ref{fig:l}). 
The surfaces for various values of $r_*$ are shown in Figure (\ref{fig:surf}).

Plugging in $g'$ given by Eq.~(\ref{eq:eom}) into the area functional Eq.~(\ref{eq:areafunc}) gives:
\begin{eqnarray}
\text{Area}(\tilde A)=2\int_0^{r_*}dr\int dy \frac{r_*^2}{(r^2-r^5)\sqrt{r_*^4-r^4}}.
\label{eq:areaon}
\end{eqnarray}
The $r$ integral diverges near $r=0$ due to the infinite volume of the near boundary AdS space. This is holographically dual to the fact that a field theory without a high energy cut-off has an infinite area law contribution to the entanglement entropy due to modes located near the boundary of the strip $A$. The area can be regulated by subtracting off the $1/r^2$ divergence of the integrand in Eq.~(\ref{eq:areaon}), which then gives a finite contribution to the area of $\tilde A$, and hence the entanglement entropy of $A$, in terms of hypergeometric functions of $r_*$. 

This allows a check of our proposed formula Eq.~(\ref{eq:heedev}) for holographic entanglement entropy: the non-relativistic field theory state dual to the black hole solution Eq.~(\ref{eq:solu}) is a finite temperature state with the same thermal properties as the black hole, given in Eq.~(\ref{eq:therm}). Therefore, the entanglement entropy of an infinitely wide strip, $l\to \infty$, should reproduce the thermal entropy of the state, up to area law contributions, which are removed by the above regulation. Examining the relation between $l$ and $r_*$, it is seen that the $l\to \infty$ limit is the same as the $r_*\to 1$ limit, and $l\approx \pi/(6\sqrt{1-r_*})$ in this regime. In this same limit, the regulated area of $\tilde A$ is:
\begin{eqnarray}
\text{Area}(\tilde A)|_\text{Reg}\approx\frac{\pi \int dy}{3\sqrt{1-r_*}}=\frac{2lL}{r_h^2},
\label{eq:areg}
\end{eqnarray} 
where we have reintroduced the length $r_h$, which was scaled out of Eq.~(\ref{eq:eom}). The prescription Eq.~(\ref{eq:heedev}) gives the entanglement entropy for this case as\footnote{Recall, this black hole solutions has $\alpha=0$.}:
\begin{eqnarray}
\lim_{l\to \infty} S^H_A|_\text{Reg}=\lim_{l\to \infty} \frac{\sqrt{1+\beta}\text{Area}(\tilde A)|_\text{Reg}}{4 G_H}=\lim_{l\to \infty} \frac{\sqrt{1+\beta}2l\int dy}{4 G_Hr_h^2}=s_{BH}\text{Area}(A),
\label{eq:entcheck}
\end{eqnarray}
where we have used that $2l\int dy$ is the area of the boundary strip $A$. The thermal entropy density $s_{BH}$ given in Eq.~(\ref{eq:therm}), as calculated from the black hole solution, is correctly reproduced by the infinite area non-relativistic entanglement entropy. Our proposal Eq.~(\ref{eq:heedev}) for non-relativistic holographic entanglement entropy therefore passes this check.

\section{Thermal nature of holographic entanglement entropy}
\label{sec:hee}

Another check of the proposed Eq.~(\ref{eq:heedev}) for holographic entanglement entropy in Ho\v rava gravity is whether a first law-like relation exists between a small subsystem's energy and entanglement entropy, defining an effective temperature, and quantifying the amount of quantum information contained in excitations \cite{Bhattacharya:2012mi}. To this end we wish to examine solutions of Ho\v rava gravity that are asymptotically holographic: beyond the Anti-de Sitter vacuum of general relativity, the Einstein-aether action Eq.~(\ref{eq:aeth}) has the $d+2$-dimensional Lifshitz vacua, for dynamical critical exponent $z$, given by:
\begin{eqnarray}
ds^2=-\frac{L^2}{r^{2z}}dt^2+\frac{L^2}{r^2}dr^2+\frac{L^2}{r^2}d\vec{x}^2,\quad u_M=\left(-\frac{L}{r^z},0,\vec{0}\right),
\label{eq:lif}
\end{eqnarray}
with $L$ a radius of curvature and the couplings $c_4=(z-1)/z$, $\Lambda=-(d+z-1)(d+z)/(2L^2)$. Due to the restricted symmetry of Ho\v rava gravity, there is an additional class of non-static vacua that have $g_{rt}=L^2/r^2$, with the coupling restricted to $\lambda= -d/(d+1)$, which is its ``conformal'' value \cite{Horava:2009uw}.

We are interested in spacetimes that asymptotically approach the above vacua at the boundary $r\to 0$. We can capture them with the ansatz:
\begin{eqnarray}
ds^2&=&-L^2\left(\frac{n(r)^2}{r^{2z}}-\frac{r^{2-2\Delta}f(r)^2n(r)^2}{g(r)}\right)dt^2+2L^2\frac{f(r)}{r^\Delta}drdt+L^2\frac{g(r)}{r^2n(r)^2}dr^2+\frac{L^2}{r^2}d\vec{x}^2,\nonumber \\
u_M&=&\left(-\frac{L n(r)}{r^z},0,\vec{0}\right),
\label{eq:pert}
\end{eqnarray}
where $n(r)=1+\mathcal{O}(r)$, $g(r)=1+\mathcal{O}(r)$, $f(r)=f_0+\mathcal{O}(r)$, and the exponent $\Delta$ is determined by the coupling $\lambda=-d(z-1)/((\Delta+d-1)(-\Delta+1+z))$. It is unclear what cases without $\Delta=2$ correspond to as they do not asymptotically approach the above vacua, but we treat them for completeness, as our generic results are independent of $\Delta$\footnote{For $\Delta<0$ the Lifshitz vacua are approached, and this ansatz is spontaneously breaking symmetry, leading to a vacuum expectation value for the operator dual to $g_{rt}$, without a corresponding source.}. Examining the equations of motion near the boundary $r\to 0$, the leading solutions are\footnote{For $\Delta \le 2$.}:
\begin{eqnarray}
n(r)\approx 1-n_{z+d}r^{z+d},\quad f(r)\approx f_0(1+n_{z+d}r^{z+d}),\quad g(r)\approx 1.
\label{eq:nbe}
\end{eqnarray}

The finite energy density of this spacetime can be calculated via a Smarr-like formula \cite{jishnu}. The Smarr ``charge'' is given by:
\begin{eqnarray}
q_{\text{Smarr}}\equiv \frac{z+1}{z+d-1}q_{\Lambda}-\frac{1+z}{2z}a_Ms^Mu_N \chi^N+\frac{K_0}{1+\beta} s_M\chi^M+\frac{\lambda K}{2(1+\beta)} s_M\chi^M,
\label{eq:qsmarr}
\end{eqnarray}
where: $q_{\Lambda}\equiv -\Lambda L^2r^d\int dr r^{-z-d-1}\sqrt{g(r)}$ is a cosmological constant contribution; $a_M\equiv u^N\nabla_N u_M$; $s_M$ is the unit spacelike vector orthogonal to $u_M$ and $K_0\equiv -u^Ms^N\nabla_N s_M$ is related to its acceleration; $\chi_M$ is the asymptotically timelike Killing vector; and $K\equiv \nabla_Mu^M$ is the trace of the extrinsic curvature of the spatial leaves. The integral of this charge over the plane spanned by $\vec{x}$ gives the total mass of the spacetime \cite{jishnu}, $M_{\text{tot}}$. Evaluating it at the boundary $r=0$ allows us to use the expansion of the metric Eq.~(\ref{eq:nbe}) which leads to an energy density:
\begin{eqnarray}
\epsilon\equiv \frac{M_{\text{tot}}}{\int d^dx}=(1+\beta)\frac{(d-z)(z+1)}{z}\frac{L^d}{8 \pi G_{H}}n_{z+d}.
\label{eq:ep}
\end{eqnarray}

We will compare the energy contained in a strip $A$ of width $l$ in one of the boundary spatial directions, $x_1$, and infinite extent in the others, with its entanglement entropy given by Eq.~(\ref{eq:heedev}). By minimizing the area of the bulk surface, $\tilde A$, anchored at the boundary of this strip, we see that it has area:
\begin{eqnarray}
\text{Area}(\tilde A)=2L^d\int_\epsilon^{r_*}dr \int d^{d-1}x \frac{1}{r^dn(r)}\sqrt{\frac{g(r)}{1-\frac{r^{2d}}{r_*^{2d}}}},
\label{eq:areastrip}
\end{eqnarray}
where $r_*$ is the deepest radius that the surface penetrates into the bulk and $\epsilon$ is a cut-off to regulate the infinite bulk volume as $r\to 0$. The turning point $r_*$ is related to the width of the strip $l$ via:
\begin{eqnarray}
l=2\int_0^{r_*} dr \frac{r^d}{r_*^d n(r)}\sqrt{\frac{g(r)}{1-\frac{r^{2d}}{r_*^{2d}}}}.
\label{eq:width2}
\end{eqnarray}

We will now assume that the width of the strip $l$, and consequently its penetration into the bulk $r_*$, are small. Quantitatively we will make the approximation $n_{z+d}l^{z+d}\ll 1$. This allows us to use the near-boundary behavior of the metric Eq.~(\ref{eq:nbe}), and expand all quantities to first order in $n_{z+d}$. This gives the area:
\begin{eqnarray}
\text{Area}(\tilde A)=\frac{2 L^d}{d-1}\int d^{d-1}x&\Bigg(&\frac{1}{\epsilon^{d-1}}-\frac{2^{d-1}\pi^{d/2}\Gamma\left(\frac{d+1}{2d}\right)^d}{l^{d-1}\Gamma\left(\frac{1}{2d}\right)^d}\nonumber\\
&+&\frac{(d-1)l^{z+1}\Gamma\left(\frac{1}{2d}\right)^{z+1}\Gamma\left(\frac{z+1}{2d}\right)n_{z+d}}{d2^{z+3}\pi^{z/2}\Gamma\left(\frac{d+1}{2d}\right)^{z+1}\Gamma\left(\frac{3d+z+1}{2d}\right)}\Bigg).
\label{eq:area2}
\end{eqnarray}
The first two terms are independent of $n_{z+d}$ and hence vacuum contributions. We will therefore use the last term as the measure of entanglement entropy that the excited state has over the vacuum. Using Eq.~(\ref{eq:heedev})\footnote{We have replaced the coupling $c_4$ with its dependence on $z$.} we obtain the ratio of entanglement entropy of the small strip to its energy:
\begin{eqnarray}
\frac{\Delta S_{ A}}{\Delta E_{ A}}=\frac{l^{z}\Gamma\left(\frac{1}{2d}\right)^{z+1}\Gamma\left(\frac{z+1}{2d}\right)}{\sqrt{1+\beta}(d-z)(z+1)d 2^{z+1}\pi^{z/2-1}\Gamma\left(\frac{d+1}{2d}\right)^{z+1}\Gamma\left(\frac{3d+z+1}{2d}\right)}.
\label{eq:efft}
\end{eqnarray}
This defines an (inverse) effective temperature by considering the entanglement entropy and energy to obey a first law of thermodynamics, $\Delta E_{ A}=T_{EE} \Delta S_{ A}$: $T_{EE}\equiv c\cdot l^{-z}$ where $c$ is a constant independent of the size of the strip, and we see the correct scaling with $l$ in terms of the dynamical critical exponent $z$, as temperature has units of inverse time\footnote{For $z\ge d$ it is unclear what the vanishing and subsequent negativity of $T_{EE}$ physically represents. It may imply an instability of such spacetimes, or that the total mass defined via Eq.~(\ref{eq:qsmarr}) is an inappropriate measure of field theory energy.}. We take this as further evidence that the proposal Eq.~(\ref{eq:heedev}) is the correct measure of entanglement entropy of a field theory region, and captures the universal nature of the amount of quantum information per energy, independent of the size of the region. 

\section{Discussion}

Entanglement entropy is a robust and useful observable for quantum theories. It is a unique order parameter, as it is applicable to phase transitions where local observables may not show critical behavior \cite{Kitaev:2005dm,2006PhRvL..96k0405L}. Condensed matter physics has many such systems, and any applicable model is a useful tool. Most of these systems have non-relativistic symmetry groups, and therefore require models further afield than typical relativistic quantum field theory. For example, Newton-Cartan geometry has recently been proposed to capture the crucial symmetries of the Quantum Hall effect \cite{Son:2013rqa}.

Non-relativistic holography is a promising tool for understanding the behavior of non-relativistic field theories. We hope that by bringing the methods of holographic entanglement entropy to Ho\v rava gravity we can obtain greater understanding of topological order in non-relativistic field theories. Using a fundamentally non-relativistic duality gives us a large class of possible well-defined holographic models; a landscape which can be explored and hopefully leads to examples of universality classes that can be realized in the lab. Furthermore, our proposal for holographic entanglement entropy from Ho\v rava gravity, Eqs.~(\ref{eq:areah}) or (\ref{eq:heeproof}), is a precise statement that can be checked versus calculations directly obtained from non-relativistic field theories. 
\\

The author would like to thank Christoph Uhlemann, Andreas Karch and Matthias Kaminski for useful feedback in preparation of the paper.

\bibliographystyle{JHEP}
\bibliography{horavaeebib}

\providecommand{\href}[2]{#2}\begingroup\raggedright\begin{thebibliography}{10}

\bibitem{Amico:2007ag}
L.~Amico, R.~Fazio, A.~Osterloh and V.~Vedral, \emph{{Entanglement in many-body
  systems}}, \href{http://dx.doi.org/10.1103/RevModPhys.80.517}{\emph{Rev. Mod.
  Phys.} {\bf 80} (2008) 517--576},
  [\href{http://arxiv.org/abs/quant-ph/0703044}{{\tt quant-ph/0703044}}].

\bibitem{Casini:2013rba}
H.~Casini, M.~Huerta and J.~A. Rosabal, \emph{{Remarks on entanglement entropy
  for gauge fields}},
  \href{http://dx.doi.org/10.1103/PhysRevD.89.085012}{\emph{Phys. Rev.} {\bf
  D89} (2014) 085012}, [\href{http://arxiv.org/abs/1312.1183}{{\tt
  1312.1183}}].

\bibitem{Kitaev:2005dm}
A.~Kitaev and J.~Preskill, \emph{{Topological entanglement entropy}},
  \href{http://dx.doi.org/10.1103/PhysRevLett.96.110404}{\emph{Phys. Rev.
  Lett.} {\bf 96} (2006) 110404},
  [\href{http://arxiv.org/abs/hep-th/0510092}{{\tt hep-th/0510092}}].

\bibitem{2006PhRvL..96k0405L}
M.~{Levin} and X.-G. {Wen}, \emph{{Detecting Topological Order in a Ground
  State Wave Function}},
  \href{http://dx.doi.org/10.1103/PhysRevLett.96.110405}{\emph{Physical Review
  Letters} {\bf 96} (Mar., 2006) 110405},
  [\href{http://arxiv.org/abs/cond-mat/0510613}{{\tt cond-mat/0510613}}].

\bibitem{Srednicki:1993im}
M.~Srednicki, \emph{{Entropy and area}},
  \href{http://dx.doi.org/10.1103/PhysRevLett.71.666}{\emph{Phys. Rev. Lett.}
  {\bf 71} (1993) 666--669}, [\href{http://arxiv.org/abs/hep-th/9303048}{{\tt
  hep-th/9303048}}].

\bibitem{Bekenstein:1973ur}
J.~D. Bekenstein, \emph{{Black holes and entropy}},
  \href{http://dx.doi.org/10.1103/PhysRevD.7.2333}{\emph{Phys. Rev.} {\bf D7}
  (1973) 2333--2346}.

\bibitem{Maldacena:1997re}
J.~M. Maldacena, \emph{{The Large N limit of superconformal field theories and
  supergravity}}, \href{http://dx.doi.org/10.1023/A:1026654312961}{\emph{Int.
  J. Theor. Phys.} {\bf 38} (1999) 1113--1133},
  [\href{http://arxiv.org/abs/hep-th/9711200}{{\tt hep-th/9711200}}].

\bibitem{Ryu:2006bv}
S.~Ryu and T.~Takayanagi, \emph{{Holographic derivation of entanglement entropy
  from AdS/CFT}},
  \href{http://dx.doi.org/10.1103/PhysRevLett.96.181602}{\emph{Phys. Rev.
  Lett.} {\bf 96} (2006) 181602},
  [\href{http://arxiv.org/abs/hep-th/0603001}{{\tt hep-th/0603001}}].

\bibitem{Nishioka:2009un}
T.~Nishioka, S.~Ryu and T.~Takayanagi, \emph{{Holographic Entanglement Entropy:
  An Overview}},
  \href{http://dx.doi.org/10.1088/1751-8113/42/50/504008}{\emph{J. Phys.} {\bf
  A42} (2009) 504008}, [\href{http://arxiv.org/abs/0905.0932}{{\tt
  0905.0932}}].

\bibitem{Janiszewski:2012nb}
S.~Janiszewski and A.~Karch, \emph{{Non-relativistic holography from Horava
  gravity}}, \href{http://dx.doi.org/10.1007/JHEP02(2013)123}{\emph{JHEP} {\bf
  02} (2013) 123}, [\href{http://arxiv.org/abs/1211.0005}{{\tt 1211.0005}}].

\bibitem{2006AnPhy.321..197S}
D.~T. {Son} and M.~{Wingate}, \emph{{General coordinate invariance and
  conformal invariance in nonrelativistic physics: Unitary Fermi gas}},
  \href{http://dx.doi.org/10.1016/j.aop.2005.11.001}{\emph{Annals of Physics}
  {\bf 321} (Jan., 2006) 197--224},
  [\href{http://arxiv.org/abs/cond-mat/0509786}{{\tt cond-mat/0509786}}].

\bibitem{Son:2008ye}
D.~Son, \emph{{Toward an AdS/cold atoms correspondence: A Geometric realization
  of the Schrodinger symmetry}},
  \href{http://dx.doi.org/10.1103/PhysRevD.78.046003}{\emph{Phys.Rev.} {\bf
  D78} (2008) 046003}, [\href{http://arxiv.org/abs/0804.3972}{{\tt
  0804.3972}}].

\bibitem{Son:2013rqa}
D.~T. Son, \emph{{Newton-Cartan Geometry and the Quantum Hall Effect}},
  \href{http://arxiv.org/abs/1306.0638}{{\tt 1306.0638}}.

\bibitem{Jensen:2014aia}
K.~Jensen, \emph{{On the coupling of Galilean-invariant field theories to
  curved spacetime}},  \href{http://arxiv.org/abs/1408.6855}{{\tt 1408.6855}}.

\bibitem{Lewkowycz:2013nqa}
A.~Lewkowycz and J.~Maldacena, \emph{{Generalized gravitational entropy}},
  \href{http://dx.doi.org/10.1007/JHEP08(2013)090}{\emph{JHEP} {\bf 08} (2013)
  090}, [\href{http://arxiv.org/abs/1304.4926}{{\tt 1304.4926}}].

\bibitem{Dong:2016hjy}
X.~Dong, A.~Lewkowycz and M.~Rangamani, \emph{{Deriving covariant holographic
  entanglement}}, \href{http://dx.doi.org/10.1007/JHEP11(2016)028}{\emph{JHEP}
  {\bf 11} (2016) 028}, [\href{http://arxiv.org/abs/1607.07506}{{\tt
  1607.07506}}].

\bibitem{Janiszewski:2014iaa}
S.~Janiszewski, \emph{{Asymptotically hyperbolic black holes in Horava
  gravity}}, \href{http://dx.doi.org/10.1007/JHEP01(2015)018}{\emph{JHEP} {\bf
  01} (2015) 018}, [\href{http://arxiv.org/abs/1401.1463}{{\tt 1401.1463}}].

\bibitem{Hubeny:2007xt}
V.~E. Hubeny, M.~Rangamani and T.~Takayanagi, \emph{{A Covariant holographic
  entanglement entropy proposal}},
  \href{http://dx.doi.org/10.1088/1126-6708/2007/07/062}{\emph{JHEP} {\bf 07}
  (2007) 062}, [\href{http://arxiv.org/abs/0705.0016}{{\tt 0705.0016}}].

\bibitem{Bhattacharya:2012mi}
J.~Bhattacharya, M.~Nozaki, T.~Takayanagi and T.~Ugajin, \emph{{Thermodynamical
  Property of Entanglement Entropy for Excited States}},
  \href{http://dx.doi.org/10.1103/PhysRevLett.110.091602}{\emph{Phys. Rev.
  Lett.} {\bf 110} (2013) 091602}, [\href{http://arxiv.org/abs/1212.1164}{{\tt
  1212.1164}}].

\bibitem{Horava:2009uw}
P.~Horava, \emph{{Quantum Gravity at a Lifshitz Point}},
  \href{http://dx.doi.org/10.1103/PhysRevD.79.084008}{\emph{Phys.Rev.} {\bf
  D79} (2009) 084008}, [\href{http://arxiv.org/abs/0901.3775}{{\tt
  0901.3775}}].

\bibitem{Jacobson:2000xp}
T.~Jacobson and D.~Mattingly, \emph{{Gravity with a dynamical preferred
  frame}}, \href{http://dx.doi.org/10.1103/PhysRevD.64.024028}{\emph{Phys.
  Rev.} {\bf D64} (2001) 024028},
  [\href{http://arxiv.org/abs/gr-qc/0007031}{{\tt gr-qc/0007031}}].

\bibitem{Barausse:2011pu}
E.~Barausse, T.~Jacobson and T.~P. Sotiriou, \emph{{Black holes in
  Einstein-aether and Horava-Lifshitz gravity}},
  \href{http://dx.doi.org/10.1103/PhysRevD.83.124043}{\emph{Phys. Rev.} {\bf
  D83} (2011) 124043}, [\href{http://arxiv.org/abs/1104.2889}{{\tt
  1104.2889}}].

\bibitem{Foster:2005ec}
B.~Z. Foster, \emph{{Metric redefinitions in Einstein-Aether theory}},
  \href{http://dx.doi.org/10.1103/PhysRevD.72.044017}{\emph{Phys. Rev.} {\bf
  D72} (2005) 044017}, [\href{http://arxiv.org/abs/gr-qc/0502066}{{\tt
  gr-qc/0502066}}].

\bibitem{Fursaev:2006ih}
D.~V. Fursaev, \emph{{Proof of the holographic formula for entanglement
  entropy}}, \href{http://dx.doi.org/10.1088/1126-6708/2006/09/018}{\emph{JHEP}
  {\bf 09} (2006) 018}, [\href{http://arxiv.org/abs/hep-th/0606184}{{\tt
  hep-th/0606184}}].

\bibitem{Germani:2009yt}
C.~Germani, A.~Kehagias and K.~Sfetsos, \emph{{Relativistic Quantum Gravity at
  a Lifshitz Point}},
  \href{http://dx.doi.org/10.1088/1126-6708/2009/09/060}{\emph{JHEP} {\bf 09}
  (2009) 060}, [\href{http://arxiv.org/abs/0906.1201}{{\tt 0906.1201}}].

\bibitem{Blas:2010hb}
D.~Blas, O.~Pujolas and S.~Sibiryakov, \emph{{Models of non-relativistic
  quantum gravity: The Good, the bad and the healthy}},
  \href{http://dx.doi.org/10.1007/JHEP04(2011)018}{\emph{JHEP} {\bf 04} (2011)
  018}, [\href{http://arxiv.org/abs/1007.3503}{{\tt 1007.3503}}].

\bibitem{jishnu}
J.~Bhattacharyya, \emph{Aspects of holography in Lorentz-violating gravity}.
\newblock PhD thesis, University of New Hampshire, 2013.

\end{thebibliography}\endgroup

\end{document}